\documentclass[preprint,preprintnumbers,amsmath,amssymb]{revtex4}
\usepackage{graphicx} % include figure files
\usepackage{dcolumn} % align table columns on decimal point
\usepackage{bm} % bold math
\usepackage{natbib}
\usepackage{hyperref}
\hypersetup{colorlinks=true,linkcolor=blue,citecolor=blue,urlcolor=blue}
\begin{document}
\title{Single Molecule Transistor based Nanopore for the detection of Nicotine}
\author{S. J. Ray}
\email{ray.sjr@gmail.com}
\affiliation{Institute of Materials Science, Technical University of Darmstadt, Alarich-Weiss-Str. 2, 64287 Darmstadt, Germany}

%\date{\today}

\begin{abstract}
A nanopore based detection methodology was proposed and investigated for the detection of Nicotine. This technique uses a Single Molecular Transistor (SMT) working as a nanopore operational in the Coulomb Blockade regime. When the Nicotine molecule is pulled through the nanopore area surrounded by the Source(S), Drain(D) and Gate electrodes, the charge stability diagram can detect the presence of the molecule and is unique for a specific molecular structure. Due to the weak coupling between the different electrodes which is set by the nanopore size, the molecular energy states stay almost unaffected by the electrostatic environment that can be realised from the charge stability diagram. Identification of different orientation and position of the Nicotine molecule within the nanopore area can be made from specific regions of overlap between different charge states on the stability diagram that could be used as an electronic fingerprint for detection. This method could be advantageous and useful to detect the presence of Nicotine in smoke which is usually performed using chemical chromatography techniques.  
\end{abstract}

\maketitle
\setcounter{figure}{0}

\section{Introduction}

Nicotine is a colourless organic compound that is often considered toxic \cite{Nicotine-toxic} when present at a significant concentration. Environmental tobacco smoke (ETS) resulting from the burning of a cigarette contains Nicotine \cite{Bratan2014} and the exposure to this is considered to be an important reason behind several chronicle diseases like lung cancer \cite{Hirayama1981,Brennan2004,Steenland1992}, asthma \cite{Otsuka2001,He1999,Pitsavos2002} and heart diseases \cite{Burchfield1996, Evans1987, Das2003}. When exposed to air, the presence of Nicotine stays significantly strong for around 2 hours and gets metabolised to Cotine at a later stage. The detection of Nicotine is thus a sensitive technique as ETS is the only possible source of Nicotine in air \cite{Bratan2014}. Usually the detection of Nicotine is performed using the techniques of radioimmunoassay \cite{Langone1973, Jacob1981}, liquid chromatography \cite{Watson1977}, gas chromatography \cite{Hartvig1979} etc. Due to the advancement in the spectrographic and chromatographic techniques, the mostly used methods use a combination of gas chromatography, mass spectrometry and high performance liquid chromatography with flame ionisation or nitrogen specific detectors \cite{Bratan2014}.  These techniques are sensitive to the amount of Nicotine and can detect a smaller presence of Nicotine in air. However, they usually require a series of chemical preparations and analysis and sophisticated instrumentations.

Here a simple but effective method was proposed for the detection of Nicotine using a nanopore based sensing technique. Conventionally a nanometer sized hole is created artificially via exposing ion beam \cite{Li2001} or electron beam \cite{Storm2003} on an electrically insulating membrane to create a solid state nanopore.  Due to the smaller size of the nanopore, when a molecule passes through a nanopore connected between two electrodes in an ionic solution, a change of the current can be detected. This detection principle is commonly used for the sequencing of DNA \cite{Clarke2009}. The materials used for constructing a nanopore are varied like Silicon \cite{Storm2003}, Silicon Nitride \cite{Li2001}, Protein \cite{Bayley2009} and Graphene \cite{Garaj2010} used in the recent times. However, the amplitude of the tunnelling current generated in the presence of a molecule within a nanopore depends on the coupling between them, which is usually small, thus limiting the detection efficiency.

By way of exploiting the weak-coupling property, a nanopore based approach has been proposed for the detection of DNA \cite{Guo2012}. Here in this work, the nanopore is essentially a single molecular transistor with the source/drain and gate electrodes placed on three sides of the nanopore. The Nicotine molecule is the active component of the device which is equivalent to the `island' in a single electron transistor. The operation of a SMT can be described using the Orthodox theory of Coulomb Blockade \cite{Grabert1992} which has been explained in details here \cite{Stokbro2010}. The usefulness of a SMT has been demonstrated for charge detection \cite{Ray2014a} and tracking the the impurity location in a single molecular device \cite{Ray2014b}.

In this case, the Nicotine molecule was placed at the centre of the nanopore and the nanopore was made in a way with one of the 3-dimensions is significantly smaller than the other two. Thus when the Nicotine molecule is pulled through the SMT, the presence of the molecule can be found from the charge stability diagram. Due to the sequential nature of transport operation, the SMT is sensitive to the orientation of the molecule and this adds additional sensitivity to the operation of the device. In this work, it was demonstrated that the charge stability diagram is unique for different orientation of the molecule and mildly sensitive to the position of the molecule within the SMT-nanopore which could be used as an electronic fingerprint for the detection of Nicotine.

\begin{figure*}
\begin{center}
\includegraphics[width=17cm]{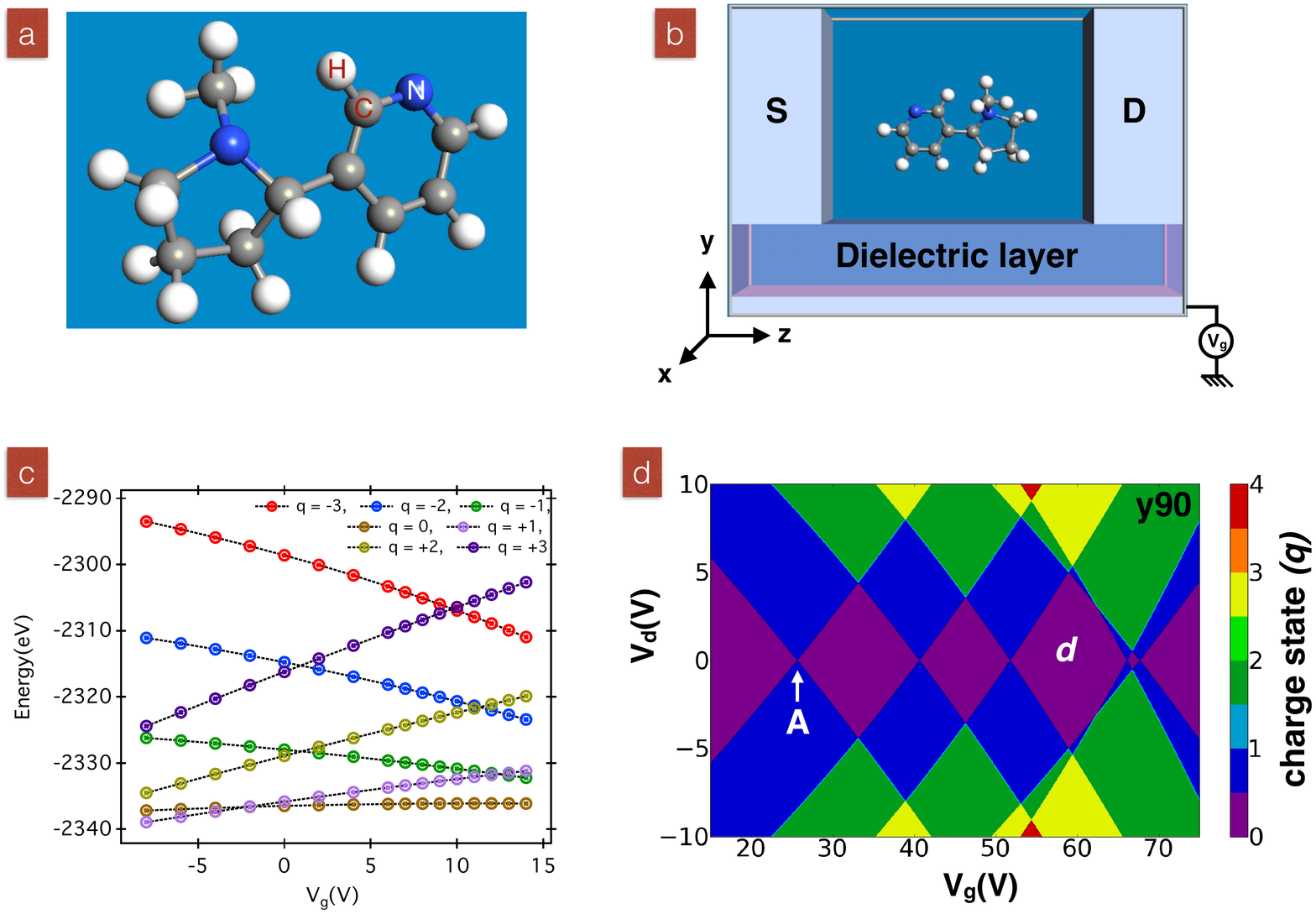}
\caption{{\small (a) Schematic of the Nicotine molecule used in the current work. The atoms in `grey', `white' and `blue' colours represent Carbon, Hydrogen and Nitrogen atoms respectively as shown in the labels, (b) A sample SMT device structure ($y90$) with the molecule placed at the centre and gate connected to the back of the dielectric layer, (c) Total energy of the molecule as function of the gate voltage for different charge states for the $y90$ configuration and (d) Charge stability diagram for the same in $y90$ configuration, the colorbar on the right represents different charge states.}}
\label{fig.1}
\end{center}
\end{figure*}

\begin{figure*}
\begin{center}
\includegraphics[width=14cm]{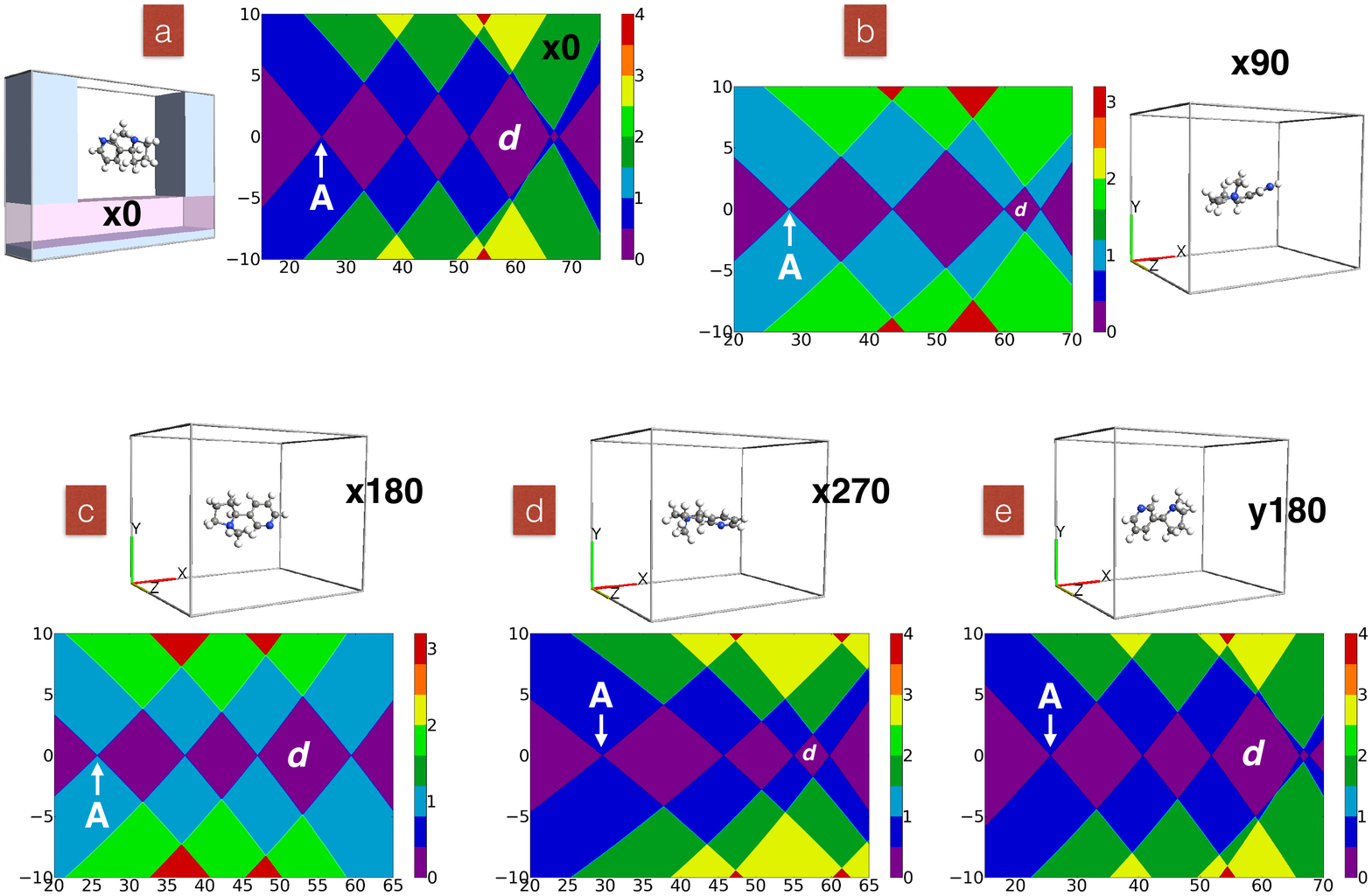} \\
\vspace{-1cm}
\includegraphics[width=14cm]{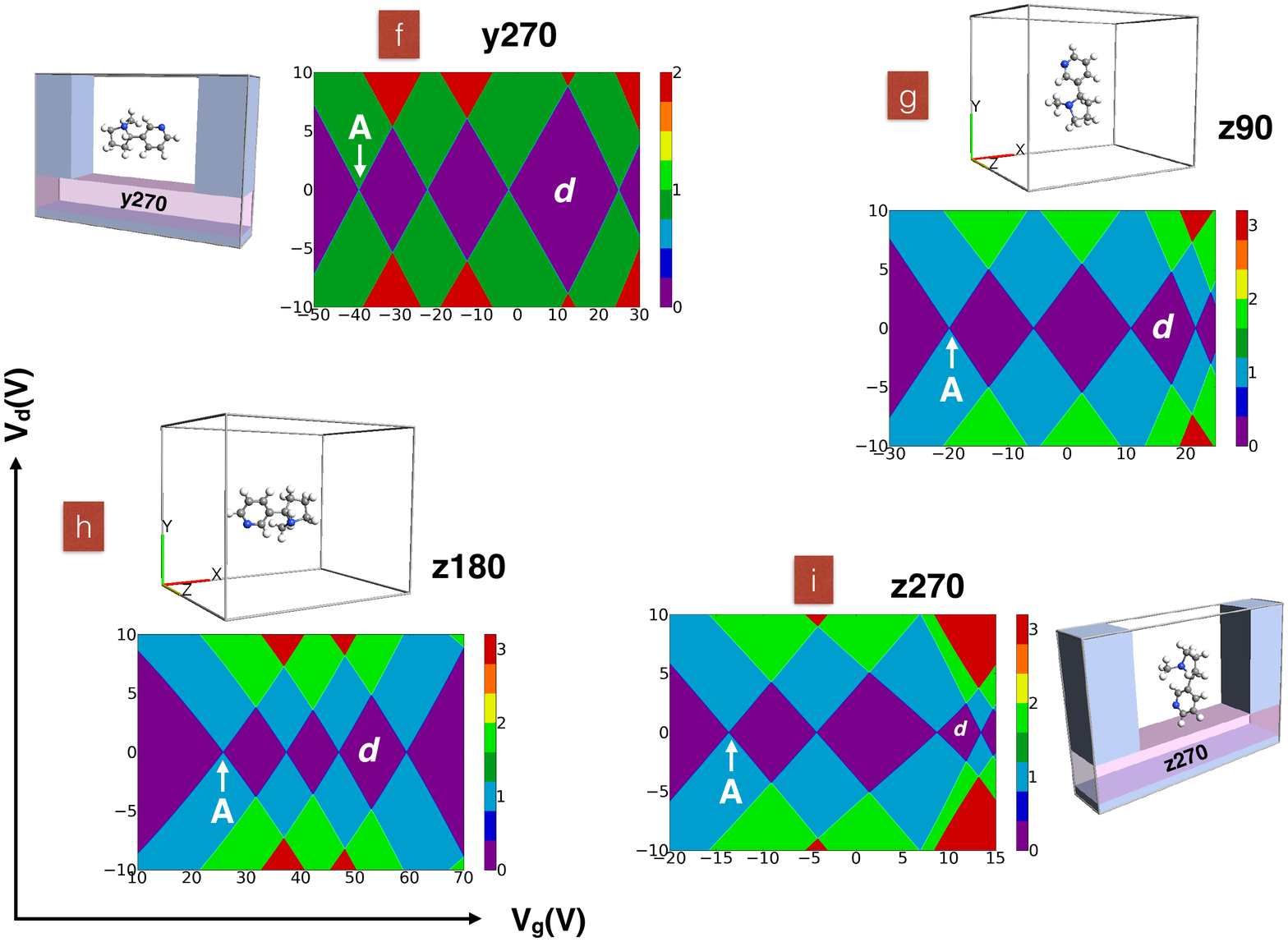} 
\caption{{\small Charge stability diagram of the molecule within the nanopore-SMT for various configurations as follows (a) $x0$, (b) $x90$, (c) $x180$, (d) $x270$, (e) $y180$, (f) $y270$, (g) $z90$, (h) $z180$ and (i) $z270$.  Schematic of the nanopore with the molecule was illustrated in (a) for $x0$, (f) for $y270$ and (i) for $z270$ configurations. The orientation of the molecule in a $xyz$-coordinate frame were shown for the other configurations in (b) $x90$, (c) $x180$, (d) $x270$, (e) $y180$, (g) $z90$ and (h) $z180$ configurations.}}
\label{fig.2}
\end{center}
\end{figure*}

\begin{figure*}
\begin{center}
\includegraphics[width=12cm]{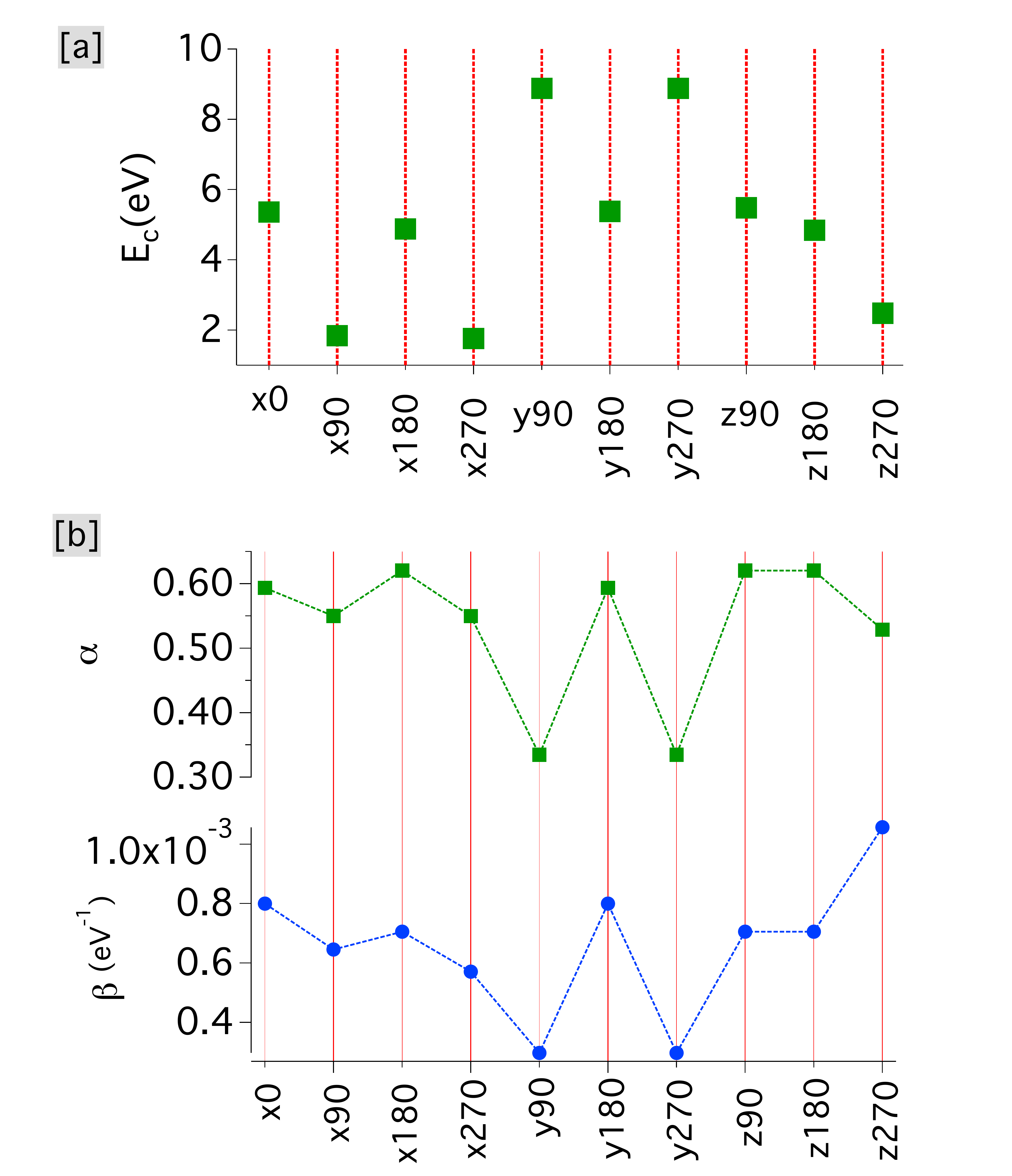}
\caption{{\small (a) Charging energy obtained from the reference diamond (marked by \emph{`d'} in Fig.~\ref{fig.1}d and Fig.~\ref{fig.2}) for different configurations of the molecule within the SMT, (b) Fitted parameters $\alpha$ and $\beta$ (from Eqn.\ref{eqn.1}) for these various configurations for the $q$=3 charge state.}}
\label{fig.3}
\end{center}
\end{figure*}

\begin{figure*}
\begin{center}
\includegraphics[width=12cm]{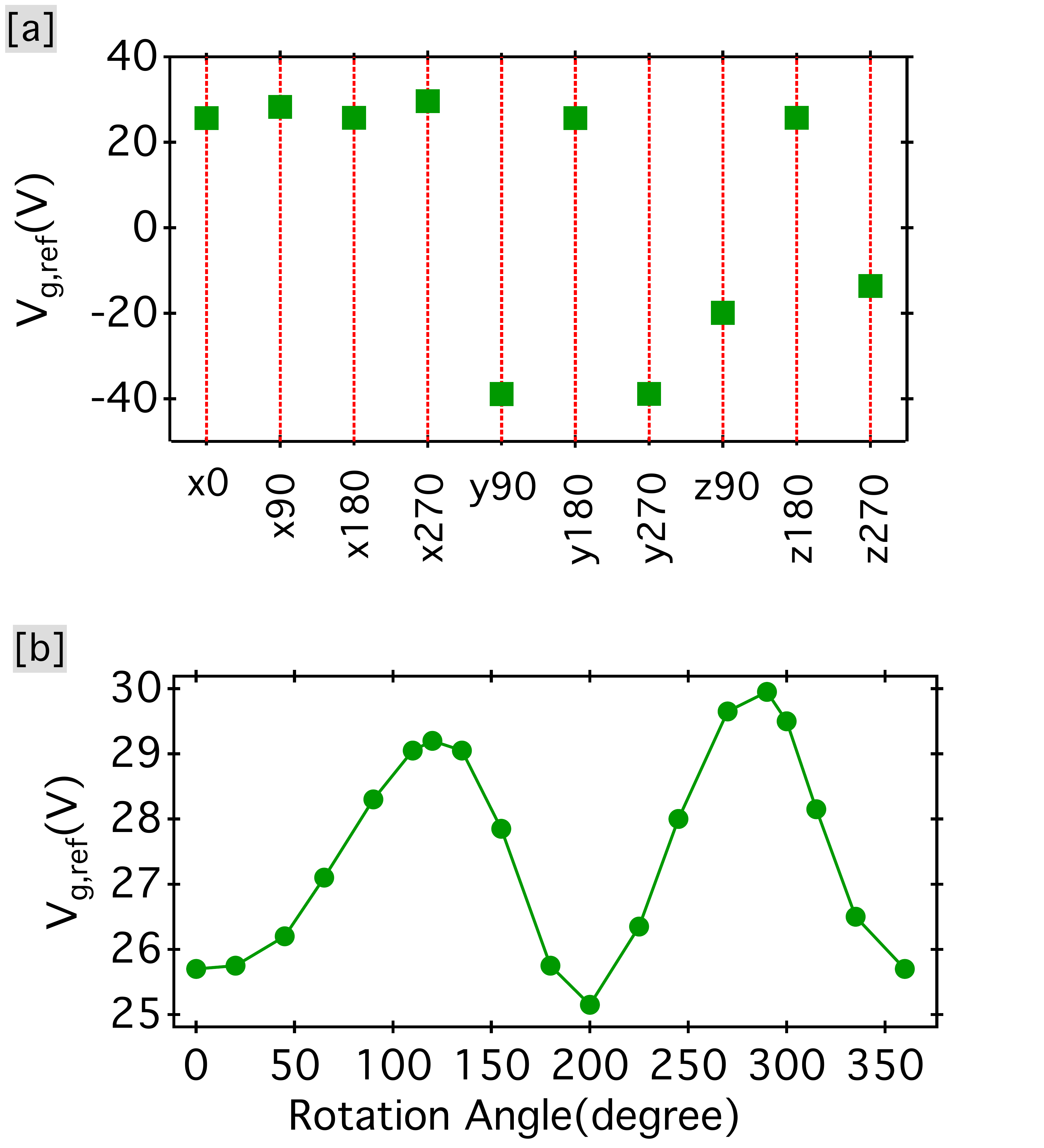}
\caption{{\small (a) V$_{\mathrm{g,ref}}$ for different configurations of the molecule as estimated from the point `A' in Fig.~\ref{fig.2}, (b) Angular dependence of V$_{\mathrm{g,ref}}$ for a continuous rotation of the molecule around the $x$-axis keeping the other coordinates fixed.}}
\label{fig.4}
\end{center}
\end{figure*}

\section{Computational Recipe}

The SMT based nanopore used in the present work has a rectangular structure which was illustrated in Fig.~\ref{fig.1}(b). The Source (S) and the Drain (D) electrodes were placed on top of a dielectric layer at its two ends. The dielectric layer has a thickness of 4 \AA\, with dielectric constant ($\varepsilon$) = 10$\varepsilon_{0}$. A metallic gate electrode of thickness 1 \AA\, was placed at the back of the dielectric layer covering the entire area. All the electrodes were considered to be made of Gold with work function ($W$) = 5.28 eV and can be replaced by other metals too. The S/D electrodes have a thickness of 5 \AA\, along the $x$-axis and the horizontal separation (along the $z$-axis) between them is 15 \AA. The height of the S/D electrodes along the $y$-axis is 12 \AA\, and the nanopore has a pore cross section of 12 \AA$\times$15 \AA\, area. Along the $x$-axis, the entire structure has a thickness of 5 \AA\, which is comparable to the Silicon \cite{Storm2003, Li2001} and Graphene\cite{Garaj2010} based nanopores reported earlier.

Nicotine is chemically known as C$_{10}$H$_{14}$N$_{2}$ with the pyridine and pyrrolidone 
rings attached to each others. This structure used for the present work was obtained using Open Babel package \cite{OpenBabel} as illustrated in Fig.~\ref{fig.1}(a). Due to the non-planar structure of the molecule with respect to the $xyz$-coordinates, the electrostatic environment is different for different orientations of the molecule within the nanopore area. Owing to the weak-coupling regime of operation, the molecular energy levels stay almost unaffected by the electrostatic environment. This 
is ensured by way of making a larger nanopore area which also allows the possibility of easier fabrication of such structure. Such a device is operational when the molecule is present within the nanopore area which can be realised from the charge stability diagram that is different from the situation when the molecule is absent within the SMT. In an experimental situation, the molecule could arrive within the nanopore in a variety of orientations of its atoms with respect to the electrodes and the corresponding response of the SMT will carry the signature of it. This was investigated systematically by way of rotating the molecule along the 3 cartesian axes and finding the influence of different molecular configurations on the charge stability diagram of the SMT.

Different configurations were named according to different rotational angles of the molecule with respect to the cartesian axes. Starting from an initial configuration (called as $x0$ as illustrated in Fig.~\ref{fig.2}(a)) \footnote{This configuration was considered to be the naturally occurring position of the molecule.}, these configurations were named following the angle of rotation from its initial state, some of which were listed in Table.~\ref{tab.1}.
\begin{table}
\begin{tabular}{|c | c |}
\hline  Configuration & Description  \\
\hline
\hline  $x0$ & Default/initial configuration.\\
\hline  $x90$ & Rotated by 90$^{\circ}$ along the $x$-axis from $x0$  \\
&  configuration keeping $(y,z)$ coordinates fixed. \\
\hline  $x180$ & Rotated by 90$^{\circ}$ along the $x$-axis from $x90$ \\
& configuration keeping $(y,z)$ coordinates fixed.\\
\hline  $y90$ & Rotated by 90$^{\circ}$ along the $y$-axis from $x0$\\
& configuration keeping $(x,z)$ coordinates fixed.\\
\hline  $z270$ & Rotated by 270$^{\circ}$ along the $z$-axis from $x0$ \\
& configuration keeping $(x,y)$ coordinates fixed.\\
\hline
\end{tabular}
\caption{Description of different device configurations}
\label{tab.1}
\end{table}%
The electrostatic behaviour of the SMT has been investigated by estimating the energy of the molecule using first-principle based calculations within a DFT framework. This computational approach was introduced by Kaajesberg $et.~al$ \cite{Kaasbjerg2008} and formulated within a DFT framework by Stokbro \cite{Stokbro2010}. The charging energy estimated by this method has been found to be in excellent agreement with the experimental results in various systems \cite{Stokbro2010}. For the present work, Ab-initio calculations were performed by using Atomistic Toolkit Package \cite{ATK} developed by the Quantum Wise division. The self consistent calculations were performed under the Local density Approximation (LDA) within a non-spin polarised DFT framework where the wave functions were expanded within a double-$\zeta$ polarised (DZP) basis set. All the metallic electrodes were considered to be equipotential surfaces and Neumann Boundary Conditions were used for solving the Poisson's equation.

\section{Results and Discussion}

The total energy of the molecule in the SMT for the $y90$ configuration was plotted as function of the gate voltage in Fig.~\ref{fig.1}(c). For different charge states, the total energy contains a fixed part contributed by the metal electrodes (=$eV$) which for the present case is 5.28 eV. Under the influence of a positive gate voltage, negative charge states have a lower energy and the reverse happens for the positive charge states at negative gate voltages. For a specific charge state $q$, the total energy varies almost linearly with the gate voltage and the dependence can be further expressed analytically by,
\begin{equation}
E(q) = E_{0} + \alpha qV_{g} + \beta (eV_{g})^2
\label{eqn.1}
\end{equation}
The first term $E_{0}$ is the zero-term in energy that is independent of the gate voltage, $e$ is the electronic charge and the proportionality constants $\alpha$ and $\beta$ could be estimated by fitting Eqn.~\ref{eqn.1} to the computationally estimated energies. For the present case, the average value of $\alpha$ has been found to be 0.3 and $\beta \sim$ 0.0013 eV$^{-1}$ from the fits as illustrated in Fig.~\ref{fig.1}(c). The value of $\alpha$ quantifies the strength of direct electrostatic coupling between the gate electrode and the molecule, while the second term is a measure of the overall polarisation of the different charge states under the electrostatic environment of the SMT. For a planar structure of the molecule lying parallel to the dielectric layer, the value of $\beta$ is usually smaller than the present case as the molecular structure is not entirely two dimensional which induces additional polarisation for different charge states. However, the value of $\alpha\gg\beta$ indicates the gate dominated linear regime of operation of the current device.

The conduction behaviour of the SMT nanopore can be understood in details from the charge stability diagram as illustrated in Fig.~\ref{fig.1}(d) for the $y90$ configuration. Due to the sequential nature of transport in the CB regime, a finite conduction in such a device only occours for certain values of the applied Source-Drain Bias ($V_{d}$) that comes within the accessible molecular energy levels. In order to observe a finite non-zero conduction, the charge population of the island has to be changed by integer numbers which can be done by changing $V_{d}$ (keeping $V_{g}$ fixed) or changing $V_{g}$ for a specific value of $V_{d}$. For a fixed value of $V_{d}$, an electron can be added or removed from the island by changing $V_{g}$. In the other way, at a specific value of $V_{g}$, the charge state of the island can be changed to an excited state via changing $V_{d}$. A series of these line scans for a certain range of the other parameter results in diamond shaped regions on the $V_{d}-V_{g}$ plane that separates the non-conducting and conducting areas of the 2D map. The large central diamond marked by `$d$' in Fig.~\ref{fig.1}(d) represents the neutral state of the molecule and the neighbouring diamonds represent the `cationic' and `anionic' state of the molecule to its `left' and `right' respectively. The charging energy of the molecule in the ground state for the neutral case can be estimated from the height of the central diamond (marked by `$d$') which is 8.88 eV for $y90$ configuration.

However, the charge stability diagram does not stay the same when the position and orientation of the molecule changes within the SMT which was illustrated in details in Fig.~\ref{fig.2}. Starting from the $x0$ configuration, the charge stability diagram changes for different configurations of the molecule when it is rotated along different Cartesian axes. Systematic evolution of the different charge states can be found from the charge stability diagram through the evolution of different coloured regions to follow the changes in different configurations. By looking at the charge stability diagrams for different configurations with naked eyes, several features could be noticed. The major changes between different configurations come from the relative orientation of the molecule (and the atoms) with respect to the electrodes which is reflected in the charging energy ($E_{c}$) behaviour of the molecule in various configurations. This can be compared for different configurations by considering a specific diamond on the charge stability diagram which for the present case has been marked by `$d$' in Fig.~\ref{fig.2}. Systematic comparison of the charging energy estimated from the height of the diamond `$d$' has been illustrated in Fig.~\ref{fig.3}(a).

It is to be noted that the energy levels on the charge stability diagram represents the charging energy of the molecule, which is the difference between two energy states [$= E(n+1) -E(n)$] and is independent of the electrical polarisation contribution towards the energy. Thus the variability of $E_{c}$ does not contain any non-linear energy term from Eqn.~\ref{eqn.1} and can be related to the effect of direct gate-molecule coupling for various configurations. From the dependence of the total energy with the gate voltage for the charge state $q$ =3, the values of the fitted parameters $\alpha$ and $\beta$ for various configurations were plotted in Fig.~\ref{fig.3}(b). For different configurations, the overall variation of $\beta$ is much smaller compared to the change in $\alpha$ and the variation of $E_{c}$ can be compared to the variation in $\alpha$.

For a given configuration of the molecule within the SMT, $\alpha$ increases when a larger volume of the molecule (or the atoms) stays in closer proximity of the dielectric layer covering the gate electrode. The difference in the values of $E_{c}$ (in Fig.~\ref{fig.3}(a)) between the $x0$ and the $x90$ configurations can be related to a similar change in the value of $\alpha$ in Fig.~\ref{fig.3}(b). By rotating the molecule further by 90$^{\circ}$ from the $x90$ configuration, the shape and size of `$d$' changes with an increase in $E_{c}$.  Due to the symmetry of the molecular structure with respect to the electrodes, $E_{c}$ for the $x0$ and $x180$ configurations are of similar values as for the $x90$ and $x270$ cases which can be found in Fig.~\ref{fig.3}(a). This can be compared to the value in $\alpha$ in Fig.~\ref{fig.3}(b) which also goes through similar variations between the $x0$, $x90$, $x180$, $x270$ configurations.  By making similar comparisons, $E_{c}$ for the $y90$ and $y270$ were estimated to be the same as for the cases of $x0$ and $y180$. However, the situation is not the same for various configurations while rotating the molecule around $z$-axis. The $E_{c}$ for $z90$ is different from $z270$ case, as in one case, the Hydrogen atoms attached to the hexagonal side faces the electrode while for the other case, the pentagonal arm comes close to the dielectric layer. The influences of gate coupling $\alpha$ is higher for $z90$ as a larger numbers of atoms come in close proximity to the gate electrode than the $z270$ case.  In Fig.~\ref{fig.3}(b), the value of $\alpha$ for $x0, x180, y180, z90$ are of similar orders which is the same for the case of $E_{c}$ for these cases in Fig.~\ref{fig.3}(a). Between the $(y90, y270)$ and $(x0, y180)$ configurations, a larger variation in $\alpha$ can be observed compared to the variations for other configurations.

In this nanopore based detection approach, the Nicotine molecule is pulled through the SMT-nanopore and the orientation of the molecule can be different from the present situation. This possibility was considered for the present case and a detection procedure was proposed that could consider these various situations. As illustrated in Fig.~\ref{fig.2}, the overall charge stability diagram changes for different configurations and specific regions of them could be considered to understand the effect of various orientations. In order to quantify this evolution in a systematic manner, a reference point was considered on the charge stability diagram which was marked by point `A' in Fig.~\ref{fig.2}. This reference point was chosen to be the left vertex of the extreme left diamond where charge states $q$ = 0 and $q$ = 1 intersects each other.  The $x$-coordinate of `A' (i.e. the value of $V_{g}$) has been plotted in Fig.~\ref{fig.4}(a) to understand the evolution of the reference point for different configurations. The differences in the values of $V_{\mathrm{g,ref}}$ for different orientations of the molecule around $x$-axis is relatively smaller than the differences in $V_{\mathrm{g,ref}}$ for similar orientations around the $y$ and $z$ axes. The value of $V_{\mathrm{g,ref}}$ for the $x0$ and $x180$ configurations are of similar order as it is the same for $x90$ and $x270$ configurations. The contrast in $V_{\mathrm{g,ref}}$ increases significantly between the $y90$ and $y180$ cases. For different configurations around the $z$-axis, the value of $V_{\mathrm{g,ref}}$ can be found to be different for the $z90$, $z180$ and $z270$ cases.

Fig.~\ref{fig.4}(a) considers only a few discrete configurations and to study further the dependence of the charge stability diagram for various rotational angles the molecule, systematic study was performed for configurations with different rotation angles around the $x$-axis keeping the other two coordinates fixed. The values of $V_{\mathrm{g,ref}}$ estimated for these cases were plotted as function of different rotation angles in Fig.~\ref{fig.4}(b). The value of $V_{\mathrm{g,ref}}$ does not overlap between two neighbouring configurations. Although a discrete set of data points were considered for this case, but the trend indicates that the value of $V_{\mathrm{g,ref}}$ is periodic for 180$^{o}$ rotation. As the value of $V_{\mathrm{g,ref}}$ is unique for different configurations, thus this could be considered as a method of identification of different configurations of the molecule within the nanopore. Since it is easy to obtain the charge stability diagram experimentally by measuring the Source-Drain current ($I_{ds}$) for different values of the $V_{d}$ and $V_{g}$, this could be used as an electronic fingerprint for the detection of Nicotine. As the SMT-nanopore works in the weak coupling limit, so the features in the charge stability diagrams are mainly determined by the energy state of the molecule. Therefore the colour pattern and the charge stability diagram carry intrinsic signatures of the presence of Nicotine molecule in a SMT-nanopore uniquely.

As Nicotine is the strongest component present in the smoke, hence a direct exposure of the smoke could be detected using this technique efficiently. However, for the case of experimental or practical realisation of such a method, several factors need to be considered. The present case only considers the situation when only Nicotine molecule is present within the SMT-nanopore area which is not necessarily the situation as the smoke also contains other elements of air.  However, the charge stability diagram is unique for each molecule and hence the presence of Nicotine could be easily identified with background subtraction by this method. In the present case, the device performance was estimated at 300K and the effect of thermal noise will be present at this temperature of operation. However, since the voltage range of operation is significantly higher ($V_{d}\sim$ 1 Volt), this contribution could be ignored. There is another possibility of fluctuation arising from the movement of the molecule itself within the nanopore. This is going to add a much smaller influence on the charge stability diagram as the weak-coupling operation of this device will have a minor influence due to this.

\section{Conclusion}

In this work, a novel approach based on a SMT based nanopore was proposed and investigated for the detection of Nicotine. The first principal based calculation was performed on such a system to obtain the charge stability diagram in the CB regime. Due to weak coupling regime of operation of this SMT-nanopore, the charge stability diagram carries unique signatures of the molecule which could be used as an electronic fingerprint for the detection of Nicotine. For various orientations of the molecule, the charge stability diagram is distinctive for different configurations that adds additional sensitivity for  identifying specific configurations of Nicotine within the nanopore area. This device becomes operational when the molecule is present within the nanopore area and the corresponding charge stability diagram is easy to obtain experimentally. Further investigation was performed by focusing on certain specific regions of the charge stability diagrams, where different charge states meet and the value of $V_{\mathrm{g,ref}}$ for these positions were found to be unique for each different configurations of the molecule within the nanopore. This methodology could work as a very powerful tool for the detection of Nicotine and Nicotine related compounds present in the smoke.

% ===================================================================== % %  Bibliography
% ===================================================================== %

\bibliographystyle{apsrev}

\end{document}